# A Domain Knowledge Informed Approach for Anomaly Detection of Electric Vehicle Interior Sounds

Deepti Kunte[a,b,∗], Bram Cornelis[a], Claudio Colangeli[a], Karl Janssens[a], Brecht Van Baelen[a] and Konstantinos Gryllias[b,c,d]

[a]*Siemens Industry Software NV, Interleuvenlaan 68, Leuven, 3001, Belgium*
[b]*KU Leuven, Department of Mechanical Engineering, Celestijnenlaan 300, Leuven, 3001, Belgium*
[c]*Flanders Make @ KU Leuven, Belgium*
[d]*Leuven.AI - KU Leuven Institute for AI, Belgium*

## ARTICLE INFO

## ABSTRACT

The detection of anomalies in automotive cabin sounds is critical for ensuring vehicle quality and maintaining passenger comfort. In many real-world settings, this task is more appropriately framed as an unsupervised learning problem rather than the supervised case due to the scarcity or complete absence of labeled faulty data. In such an unsupervised setting, the model is trained exclusively on healthy samples and detects anomalies as deviations from normal behavior. However, in the absence of labeled faulty samples for validation and the limited reliability of commonly used metrics, such as validation reconstruction error, effective model selection remains a significant challenge. To overcome these limitations, a domain-knowledge-informed approach for model selection is proposed, in which proxy-anomalies engineered through structured perturbations of healthy spectrograms are used in the validation set to support model selection. The proposed methodology is evaluated on a high-fidelity electric vehicle dataset comprising healthy and faulty cabin sounds across five representative fault types viz., Imbalance, Modulation, Whine, Wind, and Pulse Width Modulation. This dataset, generated using advanced sound synthesis techniques, and validated via expert jury assessments, has been made publicly available to facilitate further research. Experimental evaluations on the five fault cases demonstrate the selection of optimal models using proxy-anomalies, significantly outperform conventional model selection strategies.

## 1. Introduction

Sound quality in automotive cabins is a critical aspect of vehicle design, directly influencing user comfort, perceived vehicle quality, and overall brand reputation. Even minor acoustic imperfections can significantly detract from the driving experience, making the detection and mitigation of cabin sound faults a priority for automotive manufacturers.

To ensure the sound quality in vehicle cabins, end-of-line (EOL) testing has been a crucial quality assurance step, conducted just before vehicles leave the production line. While traditionally reliant on manual inspections and mechanical diagnostics, recent efforts have shifted toward sound-based and data-driven methods, offering non-invasive, scalable, and objective alternatives.

A notable industrial example is the full-vehicle NVH (Noise, Vibration, and Harshness) testing strategy developed for a high-performance hybrid vehicle, where data-driven diagnostics were applied to the front e-axle [1]. A similar direction is seen in the integration of deep learning and IoT technologies into acoustic monitoring systems on automobile production lines, enabling real-time anomaly detection and predictive maintenance [2]. Recent studies have also investigated the psychoacoustic aspects of electric vehicle interior sounds, particularly focusing on how tonal components and their subharmonics influence the perceived magnitude of tonal content (MOTC) and overall pleasantness [3; 4]. Researchers have also applied advanced vision-based anomaly detection models to an acoustic rotating machinery dataset, demonstrating the potential for real-world application in production environments [5]. Complementing these applications, the authors in [6] examined preprocessing and machine learning techniques for analyzing component operating sounds, emphasizing the importance of robust signal preparation for consistent model performance in noisy environments.

∗Corresponding author.
✉ deepti.kunte@siemens.com (D. Kunte)
ORCID(s): 0000-0001-7857-4985 (D. Kunte)

Machine learning and deep learning methods such as Support Vector Machines (SVMs), fully connected neural networks and CNNs have also been shown to be effective in classifying interior cabin noise faults such as powertrain booming noise [7–9]; Buzz, Squeak, and Rattle (BSR) [10]; Wind noise [11]; and affective sound quality characteristics [12].

Research has increasingly focused on leveraging deep learning for EOL fault detection across various subsystems. Improved detection and prediction of brake squeal was achieved using convolutional and recurrent neural networks, outperforming traditional signal-based approaches [13]. Acoustic signals have also proven effective for diagnosing electric motor faults using 1D-Convolutional Neural Network (CNN) architectures, serving as a viable alternative to vibration-based techniques in complex operational scenarios [14]. Studies have used deep learning techniques, such as MFCC-CNN and ANN, to predict and correlate transmission whine sound quality with subjective evaluations [15; 16].

Together, these industrial and academic efforts highlight a growing shift toward automated, sound-based, and machine learning-enhanced EOL testing. This transformation enables earlier fault detection, enhances diagnostic consistency, and improves the scalability of quality control processes beyond the limits of conventional methods.

Although supervised learning methods have found success in automotive fault detection scenarios, the inherent rarity of faults poses a significant challenge towards the widespread real-world adoption of these techniques. While healthy audio samples might be abundant, faulty samples are often scarce or unavailable, rendering traditional supervised learning approaches impractical. To address the scarcity of labeled fault data, anomaly detection is often formulated as an unsupervised learning problem, in which models are trained exclusively on healthy data and anomalies are identified as deviations that exceed a predefined threshold based on a chosen evaluation metric [17].

Despite their potential, unsupervised anomaly detection approaches face a fundamental challenge in model selection due to the absence of labeled faulty examples. In such settings, standard validation metrics, such as precision, recall, or F1-score, are inapplicable, as there is no ground truth against which to evaluate model performance. This makes it difficult to compare alternative models or optimize hyperparameters effectively. Furthermore, tuning hyperparameters that influence a model's sensitivity to outliers is especially problematic, as available validation feedback is limited to healthy data and does not reflect the model's ability to distinguish genuine faults from benign variations. As a result, overly complex models risk overfitting to the healthy distribution, potentially leading to poor generalization when faced with real-world anomalies, while unduly simplistic models risk underfitting, thus failing to learn the crucial intricacies of the healthy distribution. In the absence of faulty samples, model selection typically relies on surrogate metrics, such as reconstruction error or likelihood estimates, which may not reliably indicate a model's true anomaly detection capability. This highlights the need for more robust strategies for model evaluation and selection in purely unsupervised anomaly detection scenarios.

To mitigate these issues, alternative validation strategies are often employed. In some cases, researchers leverage external datasets as proxy fault data for validation, providing a basis for more informed model selection and hyperparameter tuning. One approach is the incorporation of healthy sounds from different machines or introduction of images from external datasets to simulate out-of-distribution anomalies [18; 19]. When such external datasets are not available, synthetic anomalies can be generated to approximate real-world faults [20]. Researchers have also developed augmentation strategies that manipulate healthy images by merging them with (un)modified regions from the same or different spectrograms [21–25]. These strategies offer a stand-in for genuine fault data, enabling a more robust evaluation of a model's anomaly detection capabilities and facilitating the calibration of hyperparameters. Ultimately, whether using external proxy datasets or generating engineered anomalies, these approaches lead to more reliable performance in practical applications. Nonetheless, there is a notable lack of methods specifically designed to create proxy anomalies for EOL testing scenarios involving cabin noise acoustic spectrograms.

In this paper, a novel method that leverages domain-specific knowledge of automotive acoustics to generate engineered anomalies has been proposed. The proposed approach operates directly on the spectrogram representation of cabin sound data by inducing pixel-level modifications, effectively mimicking the diverse alterations observed in real fault scenarios. The technique is both fast and straightforward to implement, enabling the rapid generation of tailored anomalies for effective model selection. Inspired by a range of problem domains, various types of changes that correspond to distinct fault conditions are captured in the perturbations introduced by the method. Empirical evaluations across five distinct cabin noise fault datasets exhibit a high correlation of the real test set with the performance of the augmented validation set constructed using the proposed engineered proxy-anomalies, resulting in optimal model selection.

While there has been commendable progress in the development of acoustic fault detection datasets, particularly for industrial and consumer applications, there remains a noticeable gap in resources specifically tailored for EOL testing in the automotive acoustics domain. Publicly available datasets such as MIMII [26; 27] and TOYADMOS [28; 29] for industrial and miniature machinery, the Electric Engine dataset [30], and the Mechanical Faults in Rotating Machinery dataset [31] have significantly contributed to advancing machine condition monitoring and fault detection. Similarly, multimodal and environmental datasets like A3CarScene [32], TICaM [33], MELAUDIS [34], and Sound of Traffic [35] have broadened the scope of audio-based analysis in driving contexts. However, these datasets are primarily focused on non-automotive or external acoustic environments and as such, offer limited relevance for benchmarking models intended for internal vehicle sound quality assessment.

Moreover, while some studies have collected or simulated datasets for cabin NVH analysis [1–4; 9–11], the majority of these datasets concentrate predominantly on internal combustion engine (ICE) vehicles, with those specifically targeting electric vehicle acoustics remaining notably scarce. Most of these datasets also remain proprietary or unpublished, primarily due to confidentiality or industrial constraints. Furthermore, many efforts rely on limited data captured under narrow conditions, which restricts their utility for broad benchmarking and model validation. This scarcity of publicly accessible, rigorously validated datasets hampers reproducibility, limits comparative evaluation of acoustic quality assessment models, and ultimately slows progress in developing robust EOL testing tools.

To address the challenge of limited and narrowly distributed fault data in automotive acoustics, a comprehensive electric vehicle cabin noise dataset that includes both healthy and faulty sound samples across five distinct fault classes has been curated. This dataset was generated using a sound quality equivalent model, built on real experimental recordings, using advanced sound synthesis techniques, ensuring that the resulting sounds accurately capture the acoustic signatures found in real-world scenarios. To validate realism of the data and establish appropriate fault levels, a jury test was conducted with domain experts in automotive acoustics. Furthermore, this dataset has been made publicly available online to foster further research and enhance reproducibility in the field [36].

In summary, this work tackles the challenge of detecting faults in car cabin sound quality under the constraint of unavailable faulty data. The key contributions of this paper are as follows:

- Development of a novel approach that leverages domain-specific knowledge to engineer proxy-anomalies in the spectrogram representation of cabin sound data. This method induces pixel-level modifications directly into the spectrogram representation, offering a fast and straightforward technique for the rapid generation of tailored anomalies to enhance model selection.

- Demonstration of significant improvements in model selection through the use of augmented validation sets, constructed with the proposed engineered proxy-anomalies. Empirical evaluations across five distinct cabin fault datasets reveal a high correlation between the performance on these augmented validation sets and the real test sets, allowing models to significantly outperform conventional model selection strategies.

- Curation of a comprehensive electric vehicle interior sound dataset featuring both healthy and faulty sound samples across five distinct fault classes, generated using a sound quality equivalent model combined with advanced sound synthesis techniques. This dataset ensures accurate capture of real-world acoustic signatures and has been validated through jury tests by experts in automotive acoustics.

- Public release of the curated electric vehicle interior sound dataset to foster further research and enhance reproducibility in the field, supporting the automotive acoustics community by providing a valuable resource for ongoing study and exploration.

The remainder of the paper is organized as follows: Section 2 provides an introduction to anomaly detection, model selection and data interpretation; Section 3 details the sound synthesis approach employed; Section 4 presents the proposed methodology for generation of engineered proxy-anomalies and the model selection algorithm; Sections 5 and 6 report experimental evaluations and comparative analyses; and Section 7 concludes with a discussion of implications and directions for future research.

## 2. Background

This section provides an overview of anomaly detection methods, focusing on anomaly detection using autoencoders. It then elaborated on model selection strategies for reconstruction error based methods. The section closes with a brief introduction to data interpretation and visualization strategies that will be used to analyze the results of this work.

## 2.1. Anomaly Detection

Anomaly detection is a crucial process that aims to identify patterns or events that significantly deviate from expected behavior [17]. Such anomalies can represent faults, defects, or unusual occurrences, potentially signaling underlying issues within complex systems. In applications like engineering and manufacturing, detecting anomalies is critical, as these often signify impending failures or quality issues, and early detection is essential for system reliability and safety.

Research in anomaly detection broadly classifies existing methods into six categories based on the underlying principles they use [17; 37; 38]:

- Classification-based techniques: These methods train models to distinguish between normal and anomalous classes. They require labeled data and include supervised approaches like decision trees, neural networks, and support vector machines. In one-class settings, they learn a boundary around normal data [39; 40].

- Nearest Neighbor-based techniques: Based on the idea that normal instances occur in dense neighborhoods, while anomalies are isolated, these methods detect anomalies by analyzing distances to nearest neighbors or local density variations [41; 42].

- Clustering-based techniques: Clustering methods assume that normal data points form large, dense clusters, while anomalies either do not belong to any cluster or form very small, sparse clusters [43].

- Statistical techniques: These methods model the distribution of data and identify points with low probability under the assumed statistical model. They can be parametric (assuming specific distributions) [44] or non-parametric [45].

- Information-Theoretic techniques: These approaches view anomalies as patterns that significantly increase the complexity or decrease the compressibility of the data. Measures like entropy or coding length are used to identify anomalous instances [46; 47].

- Spectral techniques: Spectral methods perform dimensionality reduction and detect anomalies based on deviations in the low-dimensional representation of the data [48].

Recent years have witnessed a significant shift in anomaly detection research driven by the advent of deep learning. Deep learning-based methods have demonstrated remarkable capabilities in modeling complex, high-dimensional, and structured data, enabling the discovery of subtle and previously inaccessible anomalies. Unlike traditional techniques, which often rely on predefined feature spaces and struggle with scalability or data heterogeneity, deep approaches can learn rich and expressive representations directly from raw data. This has led to substantial improvements in handling issues such as the curse of dimensionality, and the detection of conditional and group anomalies. Furthermore, the ability of deep models to generalize from limited labeled data and to integrate information from multiple data modalities has expanded the applicability of anomaly detection in diverse real-world domains [49].

### 2.1.1. Anomaly detection using autoencoders

In the realm of detecting anomalies within complex datasets, autoencoders have emerged as a powerful unsupervised learning technique. These neural network architectures are designed to efficiently encode input data into a lower-dimensional latent space and then reconstruct it, capturing essential features and minimizing reconstruction error.

Autoencoders consist of an encoder function $f_{\text{enc}}$ and a decoder function $f_{\text{dec}}$. The encoder maps the input data $\mathbf{x}$ to a latent representation $\mathbf{z}$:

$$\mathbf{z} = f_{\text{enc}}(\mathbf{x}; \theta_{\text{enc}}) \tag{1}$$

where $\theta_{\text{enc}}$ represents the encoder parameters. The decoder then reconstructs the data from this latent space back to its original form:

$$\hat{\mathbf{x}} = f_{\text{dec}}(\mathbf{z}; \theta_{\text{dec}}) \tag{2}$$

where $\theta_{\text{dec}}$ represents the decoder parameters.

The aim during training is to minimize the reconstruction error (loss), typically the mean squared error (MSE), between the input $\mathbf{x}$ and its reconstruction $\hat{\mathbf{x}}$:

$$\mathcal{L}(\mathbf{x}, \hat{\mathbf{x}}) = \frac{1}{n} \sum_{i=1}^{n} (x_i - \hat{x}_i)^2 \tag{3}$$

This loss is minimized over the dataset of normal samples, allowing the autoencoder to learn the distribution of normal data. For anomaly detection, a new sample $\mathbf{x}_j$ is fed through the autoencoder, and its reconstruction error is calculated as follows:

$$\epsilon = \mathcal{L}(\mathbf{x}_j, \hat{\mathbf{x}}_\mathbf{j}) \tag{4}$$

If the error $\epsilon$ exceeds a predefined threshold $\tau$, the sample is flagged as an anomaly:

$$\epsilon > \tau \implies \text{anomaly} \tag{5}$$

Autoencoders excel in high-dimensional anomaly detection, autonomously discerning complex data patterns without requiring labeled input or manual feature crafting. In applications like automotive acoustics, where swift and precise anomaly identification is imperative, autoencoders offer an adaptable and effective solution.

## 2.2. Model selection

Model selection and hyperparameter tuning are critical steps in the development of machine learning models, particularly in scenarios where model performance is highly sensitive to the chosen parameters. The objective of model selection is to identify the model configuration that is expected to perform optimally on an unseen test set, ensuring robustness and generalizability. Traditionally, this involves evaluating models on validation sets using metrics that are predictive of test set performance. However, in anomaly detection scenarios with only healthy data available, model selection poses unique challenges, as validation set performance cannot be directly assessed without labeled anomalies.

To address this, researchers have proposed alternative metrics for model selection in reconstruction-based anomaly detection methods. Three primary approaches are summarized below [20]:

- Reconstruction Error: Validation set reconstruction error is a commonly used metric in anomaly detection. The underlying assumption is that a model capable of accurately reconstructing or forecasting normal data should also excel at detecting anomalies. However, this metric has limitations, as it relies solely on healthy data. While it is desirable to have low reconstruction error on an unseen healthy validation set, this does not give any indication of the error on unseen faulty data. The primary objective of a classification model is to achieve a difference between reconstruction errors on healthy and faulty data samples, which cannot be evaluated with healthy validation reconstruction error alone.

- Validation Set Augmented with Synthetic Anomalies: This approach introduces synthetic anomalies into the validation set to simulate abnormal conditions. The performance of models on this augmented dataset can serve as a proxy for their anomaly detection capability. While effective, this method can suffer if the real-world anomalies differ significantly from the injected ones.

- Model Centrality: This method evaluates the consistency of outputs across different models, leveraging the idea that well-performing models tend to agree with each other. In this context, the most 'central' model among a set of trained models is considered optimal. However, this approach is imperfect, as clusters of poorly performing models may also exhibit similar outputs, misleading the centrality metric.

### 2.2.1. Hyperparameter tuning

Hyperparameter tuning plays a pivotal role in enhancing the performance of machine learning models. Even with sufficient training data, suboptimal hyperparameters can lead to underfitting, overfitting, or slow convergence. Proper hyperparameter optimization has been shown to significantly improve model generalization and stability across a wide range of machine learning applications. Below the three main approaches to tuning are discussed:

- Grid Search: Grid search systematically evaluates all possible combinations of predefined hyperparameter values [50]. While exhaustive and straightforward, this approach can be computationally expensive, especially in high-dimensional hyperparameter spaces. It is best suited for small-scale problems or when computational resources are plentiful. This technique is the sole method that guarantees optimal model selection, given a defined set of hyperparameters and a good validation metric.

- Random Search: Random search selects hyperparameter values randomly from defined ranges or distributions [51]. It is often more efficient than grid search, particularly when only a subset of hyperparameters significantly impacts model performance. By focusing on a random subset of the search space, it can achieve comparable results with fewer evaluations.

- Advanced Approaches: Advanced hyperparameter tuning methods aim to optimize the search process by intelligently exploring the hyperparameter space [52–56]. These approaches typically combine techniques to balance exploration (searching new areas of the space) and exploitation (refining promising configurations). They adapt dynamically based on previous evaluations, making them more resource-efficient than grid or random search.

### 2.3. Data interpretation and visualization

In data analysis and machine learning, various techniques, ranging from traditional statistical plots to advanced multidimensional data visualization tools, are employed for visualizing and interpreting results. In this study, t-Distributed Stochastic Neighbor Embedding (t-SNE) and saliency maps are used. These methods are particularly helpful in revealing patterns within high-dimensional data and understanding model behavior. t-SNE is utilized for dimensionality reduction and visualization of data clusters, while saliency maps are applied to enhance interpretability by highlighting features that influence model predictions.

#### 2.3.1. t-distributed stochastic neighbor embedding (t-SNE)

t-SNE is a powerful dimensionality reduction technique which is particularly effective at visualizing high-dimensional data by mapping it to a lower-dimensional space, typically two or three dimensions, while preserving the local structure of the data [57]. t-SNE works by converting high-dimensional Euclidean distances between data points into conditional probabilities that represent similarities. It then minimizes the Kullback-Leibler divergence between these probabilities and those in the lower-dimensional embedded space. This approach allows for the creation of visually interpretable embeddings that highlight clusters and structures within the data, providing insights into the underlying patterns.

#### 2.3.2. Saliency maps

Saliency maps are a technique often used in computer vision to visualize which parts of an input image are most influential in determining a model's predictions [58]. Introduced primarily for interpreting convolutional neural networks, saliency maps assign an importance score to each pixel, reflecting its contribution to the prediction. By using gradient-based methods, such as computing the partial derivative of the output concerning each input pixel, saliency maps help identify and highlight regions of interest within the data. These visualizations are crucial for model interpretability, offering a window into the decision-making process of complex models.

## 3. Dataset Generation

This section details the creation of a realistic dataset comprising both healthy and faulty vehicle interior sounds for end-of-line (EOL) testing. It begins with an overview of the data generation process from experimental recordings using virtual sound synthesis techniques. The subsequent jury test is then introduced, where perceptual assessments were used to evaluate the realism of the sounds and define thresholds for labelling the dataset. The section concludes with a summary of the final dataset.

### 3.1. Sound Quality Equivalent Modeling

In this study, sound synthesis techniques based on sound quality equivalent models [59; 60] are adopted as data generation and augmentation instrument to simulate the acoustic behavior of vehicles during EOL testing. To generate realistic interior cabin sounds, tonal and broadband components were first extracted from the experimental recordings using sound decomposition (order tracking and broadband noise tracking) techniques [61; 62].

A healthy electric vehicle sound was modeled with four dominant tonal orders, order 4, order 8, order 12, and order 16, along with a broadband noise component, as illustrated in Figure 1. Following decomposition, the tonal and broadband components were individually modified, to induce controlled randomizations to simulate variability across vehicles, and re-synthesized to generate new healthy vehicle sounds. This approach builds upon prior vehicle sound synthesis procedures proposed in [62; 63].

To extend the simulation capability to faulty scenarios, five fault types were introduced: Imbalance, Modulation, Whine, Wind, and Pulse-Width Modulation (PWM) noise (Section 3.3). Fault conditions were created by manipulation and/or addition of faulty orders and/or broadband noise components. Fault conditions also included component randomizations similar to the healthy condition. The final vehicle sounds were synthesized and made available in the time domain.

The adopted sound quality equivalent modelling approach provides a versatile and realistic means of simulating both healthy and faulty vehicle interior sounds under operational conditions, supporting the development and validation of sound-based fault detection and anomaly detection models for EOL applications.

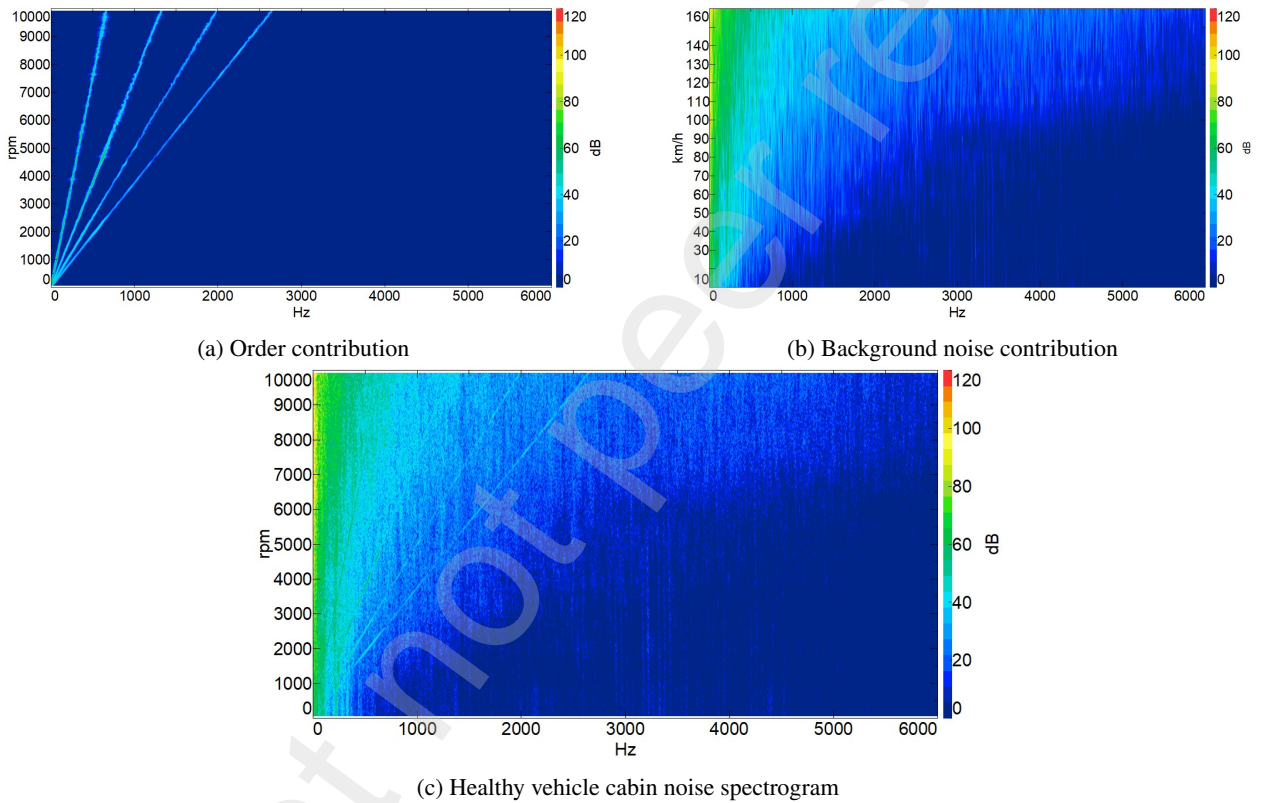

(a) Order contribution

(b) Background noise contribution

(c) Healthy vehicle cabin noise spectrogram

**Figure 1:** Orders (Figure a) and background noise (Figure b) contributors to the full vehicle interior sound. Figure c shows the vehicle's sound characteristics in healthy conditions

### 3.2. Emulation of noise implications of production line stochastic variability

In order to emulate the stochastic nature of the variations introduced by the production line and the driving test, we introduce the following variations in the sound quality equivalent model:

- Run-up profile irregularities: The EOL testing scenario considered entails a vehicle run-up. We fix the rotational speed of the motor at the start and end of the run-up at 0 and 10,000 Rotations Per Minute (RPM). This corresponds to a vehicle speed of 0 and 150 kmph respectively. While the speed values are fixed, the time it takes to complete the run-up is sampled from a uniform distribution between 15 to 25 seconds. To add some non-linearity to the speed profile, a small random error (up to 10% of its amplitude) is added to the profile mid-point, after which a cubic spline is fit to produce the resultant profile.

- **Order irregularities**: Low frequency random fluctuations (5 to 15 Hz) up to 10% of the original order amplitude were added independently to all the order profiles.

- **Background noise randomization**: Similar to order randomization, random fluctuations up to 6 dB were introduced independently on each autopower in the background noise.

### 3.3. Fault Modelling

Powertrain, road-tire interactions, wind-structure interactions, brakes and suspension are some of the major sources of NVH problems in vehicle cabins [64]. The model can be used to recreate fault scenarios by addition of faulty components and/or manipulation of healthy components. We model five distinct fault scenarios whose modelling approach is explained below:

- **Imbalance**: This scenario simulates sound variations resulting from rotational imbalances in the powertrain components of a vehicle. This is modelled by addition of two new order components, namely order 1 and order 2, to the healthy sound components. A representative sound with the Imbalance fault shown in Figure 2.

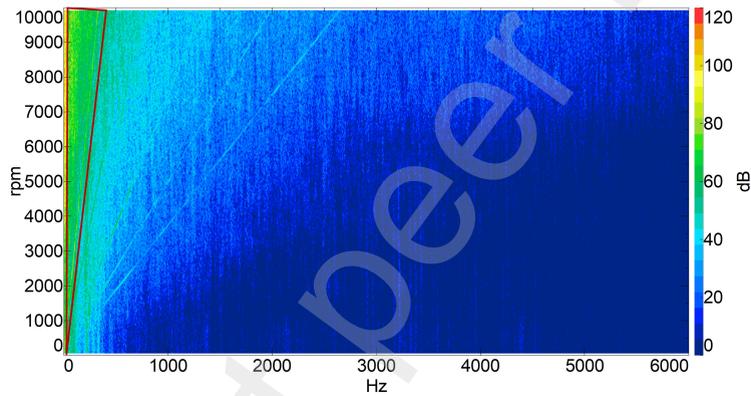

**Figure 2**: Imbalance fault vehicle cabin noise spectrogram. The region in red indicates the area affected by the fault.

- **Modulation**: Modulation occurs in presence of two or multiple closely spaced tones. Rotating machinery can generate audible modulation phenomena due interference of multiple rotating elements (elements rotating at similar, but not equal speed) and/or due to dynamic unbalance of rotors. In many cases such Modulation phenomena are undesired and in most cases an indicator of a malfunctioning of the system. In the generated dataset we emulate a Modulation phenomenon by adding 0.1 order-spaced sidebands to order 8 (i.e. order 7.9 and 8.1). The added orders have similar amplitude profile compared to order 8. Their amplitude levels can be amplified or attenuated to emulate severe and mild Modulation phenomena respectively. A unit-amplitude Hanning window was applied to the faulty orders to smoothen the introduced curves. To introduce further variability, the start and end points for the window were sampled from a uniform distribution within [2500, 3500] and [6500, 7500] RPM ranges, respectively. This introduced variability in the placement as well as width of the fault, allowing it to range from as compact as 3000 RPM to as wide as 5000 RPM at the extremes. A representative sound with the Modulation fault shown in Figure 3.

- **Whine**: Whine noise, high-pitched and prominent tonal sound, in automotive is typically caused by gearbox phenomena [65] or e-motor-related phenomena. Other factors influencing whine are variation of meshing stiffness, dynamic meshing forces, friction forces and ineffective lubrication [66]. The whine fault was introduced solely by manipulating a healthy order component (order 16). The healthy order 16 was modified by multiplication with a hanning window. The start and end points for the window were sampled from a uniform distribution within [5500, 6500] and [7500, 8500] RPM ranges, respectively. A representative sound with the Whine fault is shown in Figure 4.

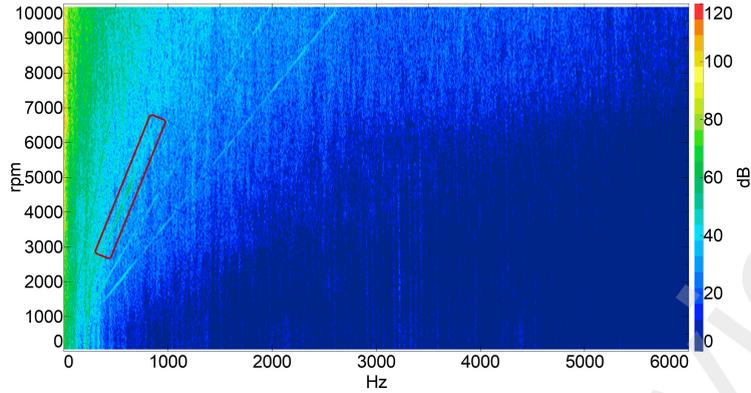

**Figure 3:** Modulation fault vehicle cabin noise spectrogram. The region in red indicates the area affected by the fault.

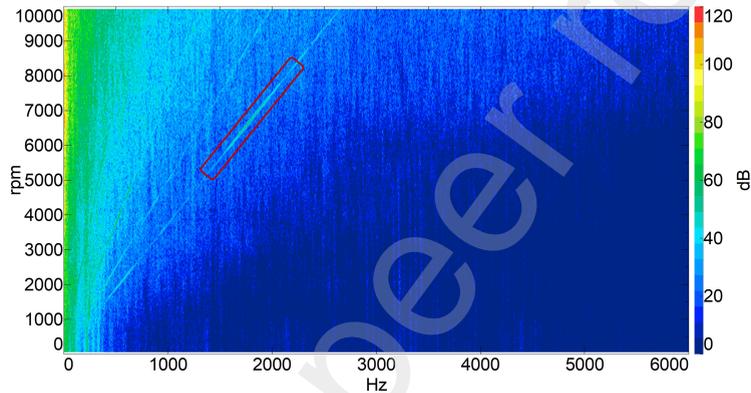

**Figure 4:** Whine fault vehicle cabin noise spectrogram. The region in red indicates the area affected by the fault.

- **Wind**: Wind noise is a major contributor to a vehicle sound. In production, this noise component can diverge from design targets when either flush and gaps cause undesired exterior whistle sound or, like in the case at hand, when air leaks inside the vehicle compartment due to suboptimal doors sealing in operation. When leakage occurs, it results in a band-limited broadband blowing sound. The phenomenon is directly proportional to vehicle speed [11; 67]. This fault was modelled by the addition of a faulty broadband noise component. The frequency range of the faulty component was limited by multiplication with a unit-amplitude hanning window. The start and end points for the window were sampled from a uniform distribution within [3750, 4250] and [5250, 5750] Hz frequency ranges, respectively. A representative sound with the Wind noise fault is shown in Figure 5.

- **Pulse Width Modulation (PWM)**: In electric vehicle drive systems, electromagnetic interference (EMI) noise is produced due to voltage fluctuations occurring during switching operations performed in the inverter and battery circuits [68]. This scenario was modelled via additional $0^{th}$, $24^{th}$, and $-24^{th}$ orders at 5000 Hz. A representative sound with the PWM fault shown in Figure 6.

Another dimension of variation for the faults was the fault amplitude. Faults were adjusted on a dB scale by either scaling the introduced faulty components (for Imbalance, Modulation, Wind, and PWM faults) or by modifying the Hanning window multiplier (for Whine fault). To determine a realistic range of fault amplitudes for dataset generation, a jury test was conducted, as described in the following subsection. The results, including the audible fault amplitudes and the amplitude ranges used for data generation, are summarized in Tables 1 and 2, respectively.

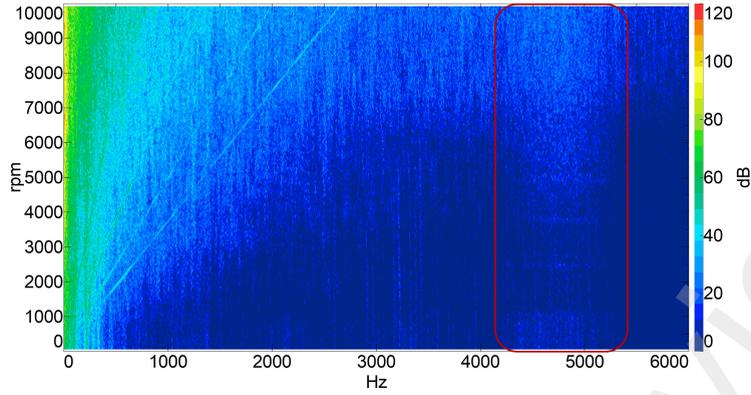

**Figure 5**: Wind noise fault vehicle cabin noise spectrogram. The region in red indicates the area affected by the fault.

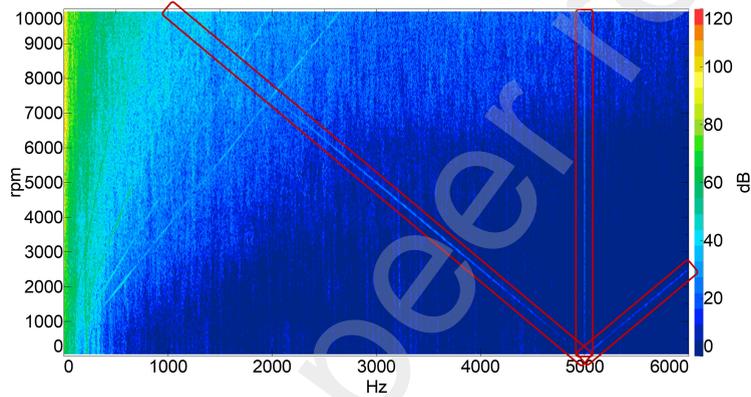

**Figure 6**: PWM fault vehicle cabin noise spectrogram. The region in red indicates the area affected by the fault.

### 3.4. Jury Testing

To validate realism of the generated sounds and plausibility of the introduced faults, a jury test was conducted with sound quality experts. The test aimed to assess two key aspects: (1) whether the generated sounds were perceived as realistic by the experts and (2) the amplitude threshold at which each fault type becomes audible. This feedback guided adjustments to the sound model to ensure realistic and perceptually accurate fault representation.

All jurors used high-fidelity headphones and a sound splitter box to ensure consistent sound quality and playback for each participant. This setup ensured that all jurors experienced the sounds under identical conditions, minimizing variability and enhancing the reliability of their assessments.

#### 3.4.1. Test Procedure

The jury test was conducted in two parts:

- Part 1: Realism assessment
  In the first part of the test, jurors were presented with a series of healthy and faulty sounds, with varying levels of background noise. The objective was to assess whether the generated sounds were realistic. The jurors agreed that the sounds indeed sounded realistic, and the different faults were accurately represented. They also gave feedback on the amount of acceptable randomization that could be added to the different sound components.

- Part 2: Fault audibility threshold setting
  In the second part of the test, the focus shifted to identifying the fault audibility threshold. A series of sounds with varying fault amplitudes were generated for evaluation. Jurors listened to these sounds simultaneously and rated their audibility.

To ensure thorough assessments, they could replay the current sound or any previous sound files as needed. The objective was to reach a consensus on whether each fault was audible, categorizing their evaluations as *Yes*, *Maybe*, or *No*.

#### 3.4.2. Jury Test Outcome

In both parts of the test, the jurors reached a consensus, providing consistent judgments regarding sound realism and fault audibility across all faults. The results of the test are summarized in the following table, which categorizes the detection of various fault types by the jurors:

Table 1 presents the jurors' assessments of various fault types, indicating the amplitude levels (in dB) at which each fault was perceived. For each fault, the *No* category specifies amplitude levels where the fault was completely inaudible. The *Yes* levels represent amplitudes at which the fault was clearly audible. In contrast, the *Maybe* levels denote amplitudes where there may be a slight indication of a fault, without sufficient clarity and consensus to definitively classify it as faulty.

The results reveal significant variability in detection thresholds for different faults, suggesting that some faults are easier to perceive than others. For instance, the Modulation fault was audible at much lower amplitudes compared to the Whine, which only became noticeable at higher amplitudes.

Additionally, it is noteworthy that the *Maybe* regions vary in complexity; for example, the Modulation and Whine faults each have only one amplitude level in the *Maybe* category, while the Wind fault includes three levels. This variability emphasizes the complexity of sound perception and highlights how sensitivities to different faults can vary significantly. Such insights are crucial for understanding the nuances of automotive cabin acoustics, ultimately aiding in the development of more realistic sound quality evaluations.

### 3.5. Dataset creation

Having determined the minimum audible amplitude of the faults, this was used as the lower extreme when scaling them. The maximum amplitude of the scaling range was kept 12 dB above the minimum giving a wide range of amplitude values for the faults. Table 2 shows the fault amplitudes introduced as well as the number of sounds created per category to generate the anomaly detection dataset.

## 4. Methodology

This section details the methodology for detecting anomalies in vehicle cabin sounds along with the proposed model selection approach. It begins with an overview of the data pre-processing, which transforms raw audio into spectrograms for anomaly detection. Next, the generation of proxy-anomalies is discussed, where controlled perturbations are applied to healthy validation data to simulate various fault types. The section concludes with a description of the anomaly detection model and the model selection process using validation data augmented with the generated proxy-anomalies.

### 4.1. Pre-processing

The data preprocessing begins by converting raw audio files into spectrograms, which offer a time-frequency view of the audio signals. Spectrograms are created by applying a series of Fourier transforms at different RPMs or distinct time intervals during machinery operation, resulting in a three-dimensional representation with frequency, RPM (or time), and amplitude as axes. This transformation is particularly useful, as it reduces the data's dimensionality while retaining essential information, which aids machine learning models in detecting patterns for predictive tasks. An added advantage of employing order-tracked spectrograms is the resulting uniformity in input representation. Since the speed range can be precisely selected, this method ensures that the spectrogram dimensions remain consistent, irrespective of variations in the duration of the sound data. This is particularly useful when using machine learning models. After generating the spectrograms, the dataset clipped to the range of 0 to 120 dB and then linearly scaled to fit in the interval [0, 1]. Rather than normalizing each spectrogram individually, this approach maintains uniform scaling, contributing to consistent data representation throughout the dataset.

The area, i.e., number of pixels in the spectrogram occupied by the different faults, will now be examined. The spectrogram itself has dimensions of 1536 x 384 pixels, representing frequency and RPM axes of the signal respectively. As shown in Table 3, the wind fault, which is broadband in nature and is present in a wide frequency range, alters a significant portion of the spectrogram. On the other hand, the other faults, which are order-based affect significantly smaller areas. Particularly the whine and modulation faults, which affect only partial order regions, are seen in the least number of pixels.

**Table 1**
Fault audibility thresholds in dB for different fault classes

| Fault Type | No | Maybe | Yes |
|---|---|---|---|
| Imbalance | ...-5, -4 | -3, -2 | -1, 0,... |
| Modulation | ...-14, -13 | -12 | -11, -10,... |
| Whine | ...9, 10 | 11 | 12, 13,... |
| PWM | ...8, 9 | 10, 11 | 12, 13,... |
| Wind | ...-12, -11 | -10, -9, -8 | -7, -6,... |

**Table 2**
Minimum and Maximum Amplitudes for Fault Detection and Number of Samples

| Class | Minimum Amplitude | Maximum Amplitude | Number of Sound Samples |
|---|---|---|---|
| Healthy | - | - | 1200 |
| Imbalance fault | -1 | 11 | 200 |
| Modulation fault | -11 | 1 | 200 |
| Whine fault | 12 | 24 | 200 |
| PWM fault | 12 | 24 | 200 |
| Wind noise fault | -7 | 5 | 200 |

**Table 3**
Fault prevalence in the input representation

| Fault | Number of pixels affected | % area affected |
|---|---|---|
| Imbalance | 5399 | 0.92 |
| Modulation | 1409 to 2340 | 0.24 to 0.4 |
| Whine | 314 to 628 | 0.05 to 0.16 |
| PWM | 6409 | 1.09 |
| Wind | 96062 to 192125 | 16.29 to 32.57 |

### 4.2. Proxy-Anomaly Generation

To enable model selection without relying on real anomalous data, anomalies are synthetically introduced by perturbing healthy validation data, creating a controlled set of anomalous validation data. The engineered anomalies require minimal computational resources are are thus time-efficient. Meanwhile, they are produced through domain knowledge inspired perturbations, reflecting potential faults commonly observed in vehicle cabin noise. This enables the model selection process to prioritize models that can generalize to detect anomalies across a broad spectrum of potential real-world faults, ensuring robust performance under diverse conditions.

To create proxy-anomalies for model selection, three types of structured perturbations are employed viz. *add_rpm*, *add_freq*, and *add_order*. Each perturbation targets specific audio characteristics relevant to vehicle cabin noise such as impacts, tonal noise and order fluctuations. All perturbations are applied on healthy validation data samples to synthetically produce proxies for anomalous data samples. Below, the synthetic perturbations are formulated, after which the parameters introduced in their definition are listed (Table 4).

- Addition of RPM-Lines (*add_rpm*): The *add_rpm* perturbation adds a specific randomly selected amplitude to pixels along multiple adjacent RPM lines in the spectrogram. This is done by selecting $n_{rpm}$ RPM lines, starting from a randomly chosen RPM index $r$, and adding a consistent amplitude value $\delta_{\text{rpm}}$ across all pixels within these lines. The perturbed spectrogram $S_{\text{rpm}}$ is defined as:

$$S_{\text{rpm}}(i,j) = \begin{cases} S(i,j) + \delta_{\text{rpm}} & \text{if } r \leq i < r + n_{rpm} \\ S(i,j) & \text{otherwise} \end{cases} \quad (6)$$

where $\delta_{\mathrm{rpm}}$ is a random amplitude within the range $[a_{rpm}, b_{rpm}]$, and $n_{rpm}$ specifies the width of the perturbed RPM region. Index $r$ is sampled from the entire RPM range of the given spectrogram. The effect of the perturbation on the spectrogram is shown by an example (c.f. Figure 7).

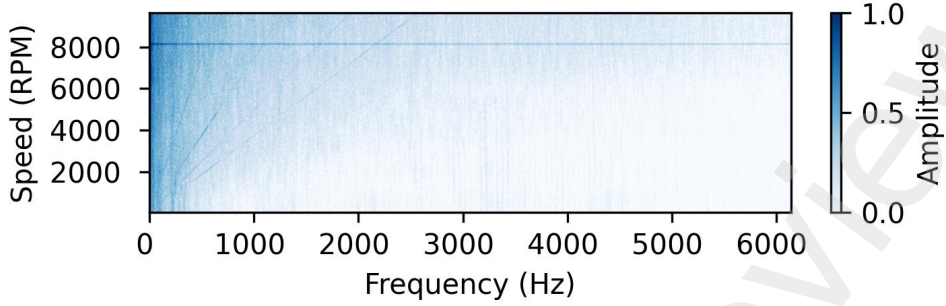

**Figure 7:** Synthetic anomaly: Addition of RPM line

- Addition of Frequency-Lines (add_freq): The *add_freq* perturbation works in a similar way to *add_rpm*. Instead of targeting RPM lines, it targets adjacent frequency lines. The resulting perturbed spectrogram $S_{\mathrm{freq}}$ is given by equation 7

$$S_{\mathrm{freq}}(i,j) = \begin{cases} S(i,j) + \delta_{\mathrm{freq}} & \text{if } f \leq j < f + n_{freq} \\ S(i,j) & \text{otherwise} \end{cases} \quad (7)$$

where $\delta_{\mathrm{freq}}$ is selected from the interval $[a_{freq}, b_{freq}]$, and $n_{freq}$ represents the number of consecutive frequency lines altered. Index $f$ is sampled from the entire frequency range of the given spectrogram. Figure 8 presents a spectrogram with the applied perturbation.

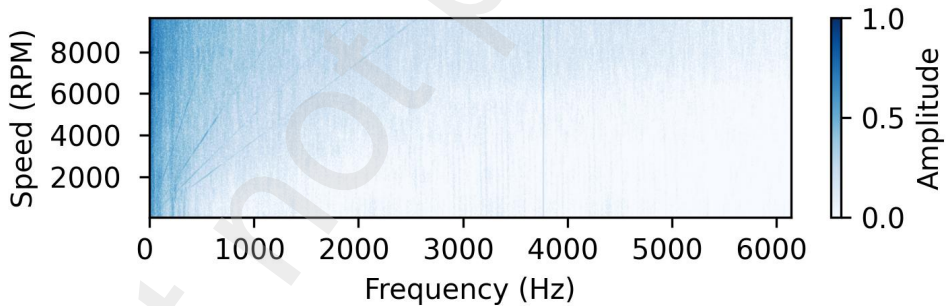

**Figure 8:** Synthetic anomaly: Addition of frequency line

- Addition of Order-Lines (add_order): Similar to the *add_rpm* and *add_freq* perturbations, *add_order* modifies specific order lines within the spectrogram by adding a consistent amplitude offset. The resulting perturbed spectrogram $S_{\mathrm{order}}$ is given by equation 8:

$$S_{\mathrm{order}}(i,j) = \begin{cases} S(i,j) + \delta_{\mathrm{order}} & \text{if } i - t/2 \leq j * n_{order} * c < i + t/2 \\ S(i,j) & \text{otherwise} \end{cases} \quad (8)$$

where $\delta_{\mathrm{order}}$ is randomly chosen from the interval $[a_{\mathrm{order}}, b_{\mathrm{order}}]$, and $t$ defines the size of the region modified. The order region to be altered is randomly chosen from a specified set of orders $[c_{\mathrm{order}}, c_{\mathrm{order}} + 0.5, c_{\mathrm{order}} + 1.0, c_{\mathrm{order}} + 1.5, ..., d_{\mathrm{order}}]$, i.e. all half-orders in the interval $[c_{\mathrm{order}}, d_{\mathrm{order}}]$. A spectrogram modified by this perturbation is depicted in Figure 9.

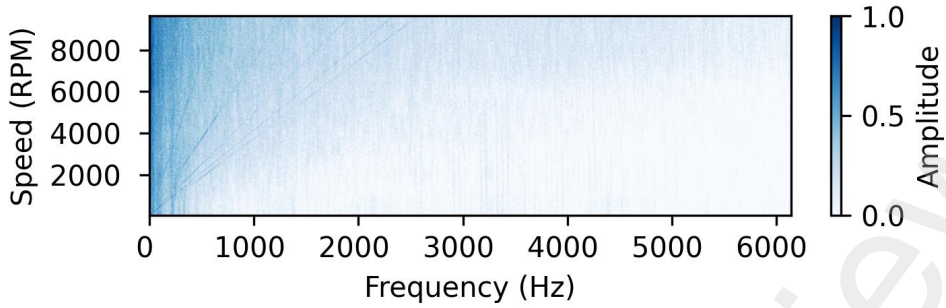

**Figure 9:** Synthetic anomaly: Addition of order line

**Table 4**
Parameters used during generation of proxy-anomalies

| Perturbation Type | Parameter | Value |
|---|---|---|
| add_rpm | $a_{\text{rpm}}$ | 0.1 |
| | $b_{\text{rpm}}$ | 0.25 |
| | $n_{\text{rpm}}$ | 4 |
| add_freq | $a_{\text{freq}}$ | 0.1 |
| | $b_{\text{freq}}$ | 0.25 |
| | $n_{\text{freq}}$ | 4 |
| add_order | $a_{\text{order}}$ | 0.1 |
| | $b_{\text{order}}$ | 0.25 |
| | $t_{\text{order}}$ | 15 Hz |
| | $c_{\text{order}}$ | 2 |
| | $d_{\text{order}}$ | 40 |

### 4.3. Anomaly Detection Model

To demonstrate the model selection capabilities in anomaly detection, a Convolutional Neural Network (CNN) autoencoder architecture, which is well-suited for extracting spatial hierarchies and patterns from spectrogram data has been used. The autoencoder model consists of an encoder that compresses input spectrograms into a compact latent representation and a decoder that reconstructs the original spectrogram from this latent space.

The encoder processes each spectrogram through a series of convolutional layers, progressively reducing the spatial dimensions while increasing the depth of feature maps. Each convolutional layer is followed by activation functions to introduce non-linearity and by pooling layers to further downsample the feature maps, preserving critical features while reducing the data dimensions.

The decoder reconstructs the spectrogram by successively upsampling the latent vector using transpose convolutional layers, which increase the spatial dimensions back to the original size. Through this process, the decoder aims to reconstruct the normal spectrogram as closely as possible, allowing the model to learn the underlying structure of healthy data. Any significant deviations in reconstruction error between the input and output spectrogram can serve as an indicator of potential anomalies.

While this study demonstrates the applicability using a CNN architecture, it is noteworthy that the approach is not constrained by a specific architecture and can similarly be employed on more advanced architectures. The primary focus of this paper is the concept of model selection in autoencoder-based anomaly detection rather than the architecture itself.

### 4.4. Model selection

The proposed model selection strategy is detailed in Algorithm 1. Unlike conventional approaches that rely solely on reconstruction error computed from a healthy validation set, the proposed method evaluates model performance on an augmented validation set comprising both healthy samples and proxy anomalies. This shift enables a more representative assessment of the model's anomaly detection capability during the selection process. The training and model selection process is detailed below:

- Training: The model is trained exclusively on healthy training samples. No faulty data is used at this stage.

- Early stopping: Healthy validation samples are used to monitor the reconstruction loss and determine the optimal stopping point during training.

- Proxy-anomaly generation: After training, proxy-anomalies are generated by perturbing the healthy validation samples using domain-specific transformations. These proxy anomalies, combined with the original healthy validation data, form an augmented validation set.

- Model selection: The trained model is evaluated on the augmented validation set. The performance on detecting proxy-anomalies is used to select the final model. Performance evaluation has been done using Area Under Receiver Operating Characteristic (AUROC) curve.

---

Algorithm 1: Model selection using proxy-anomalies

---

**Input:** Healthy dataset $\mathcal{D}^H$, candidate architectures $\{\mathcal{A}_1, \ldots, \mathcal{A}_N\}$, proxy-anomaly perturbation $\{P_1, P_2, \ldots, P_p\}$, number of runs $R$

**Output:** Selected architecture $\mathcal{A}^*$

**for** each architecture $\mathcal{A}_i$ in $\{\mathcal{A}_1, \ldots, \mathcal{A}_N\}$ **do**

    **for** $r = 1$ to $R$ **do**

        Split healthy dataset $\mathcal{D}^H$ into training set $\mathcal{D}^H_{\text{train}}$ and validation set $\mathcal{D}^H_{\text{val}}$ randomly

        Initialize model $\mathcal{M}_i$ with architecture $\mathcal{A}_i$

        Train model $\mathcal{M}_i$ on training set $\mathcal{D}^H_{\text{train}}$

        Use $\mathcal{D}^H_{\text{val}}$ to apply early stopping based on reconstruction loss

        Monitor reconstruction loss on $\mathcal{D}^H_{\text{val}}$ for early stopping

        Divide the healthy validation set $\mathcal{D}^H_{\text{val}}$ into $p$ subsets based on the number of perturbation types

        **for** each proxy-anomaly perturbation type $P_i \in \{P_1, P_2, \ldots, P_p\}$ **do**

            For each $x_k \in \mathcal{D}^{H_p}_{\text{val}}$, generate proxy-anomaly $\tilde{x}_k = \text{Perturb}_i(x_k)$ using perturbation $P_i$

            Assign the generated anomalies to $\mathcal{D}^{P_i}_{\text{val}}$

        **end for**

        Construct the proxy-anomaly validation set $\mathcal{D}^P_{\text{val}} = \mathcal{D}^{P_1}_{\text{val}} \cup \mathcal{D}^{P_2}_{\text{val}} \cup \cdots \cup \mathcal{D}^{P_p}_{\text{val}}$

        Construct augmented validation set $\mathcal{D}^{\text{aug}}_{\text{val}} = \mathcal{D}^H_{\text{val}} \cup \mathcal{D}^P_{\text{val}}$

        Evaluate model $\mathcal{M}_i$ on $\mathcal{D}^{\text{aug}}_{\text{val}}$ using AUROC

        Store AUROC score for run $r$, $s_{i,r}$

    **end for**

    $s_i \leftarrow \frac{1}{R} \sum_{r=1}^{R} s_{i,r}$

**end for**

Select architecture $\mathcal{A}^* = \arg\max_{\mathcal{A}_i}(s_i)$ where $s_i$ is the average AUROC score over $R$ runs

---

## 5. Experiments

In this study, the impact of varying model architectures on performance has been explored by manipulating two key parameters: the number of blocks ($n$) and the number of filters ($f$) in the convolutional layers (c.f Figure 10). The choice of these hyperparameters is based on their fundamental impact on the network's complexity and thus its learning capacity. The number of blocks ($n$) ranged from 1 to 7 while number of filters could be either 4, 8, 16, or 32. This resulted in a comprehensive evaluation of 28 distinct model configurations, each designed to assess the effect of increased complexity and learning capacity on the model's ability to accurately reconstruct and represent the input data. For each configuration, the models were trained on a fixed dataset, while monitoring the healthy validation set reconstruction error. The fixed hyperparameters of the networks are listed in Table 5 and the model architecture is shown in Figure 10 respectively.

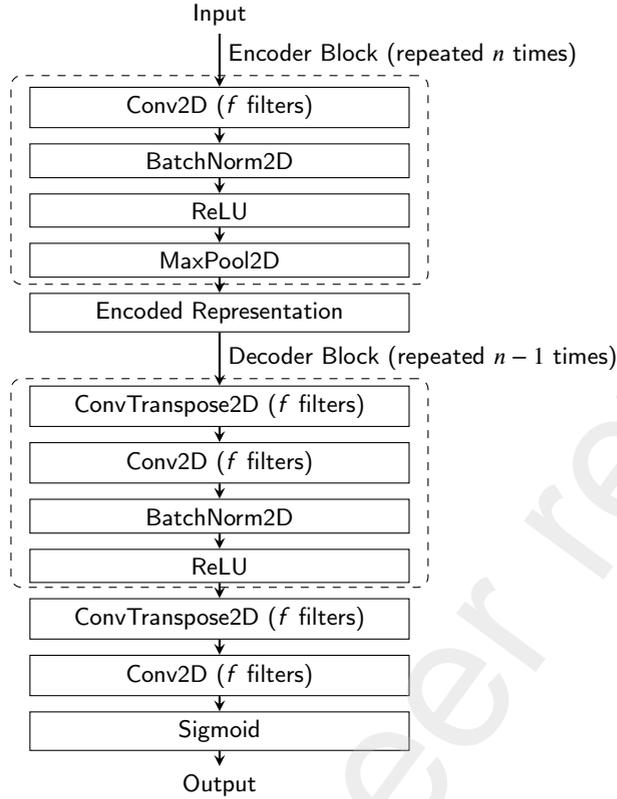

**Figure 10:** Autoencoder model architecture for anomaly detection. Dotted lines indicate the encoder and decoder blocks to be repeated based on the selected hyperparameter

**Table 5**
Fixed hyperparameters of the Autoencoder metwork

| Hyperparameter | Value |
| --- | --- |
| Learning rate | 0.0001 |
| Batch size | 8 |
| Maximum number of epochs | 100 |
| Optimizer | Adam |
| Loss Function | Mean squared error |
| Convolutional layer kernel size | 3 |
| Convolutional layer stride | 1 |
| Convolutional layer padding | 1 |
| Pooling layer type | MaxPool |
| Pooling layer kernel size | 2 |
| Pooling layer stride | 2 |

In order to evaluate the effectiveness of the proposed model selection strategies, we establish two baseline methods commonly considered in anomaly detection. Additionally, we also establish a best case scenario. While the ideal case uses information that is not available during training and model selection, it represents the performance target that the model selection strategy should aim to achieve. Furthermore, we also compare the proxy-anomaly based model selection to model selection using other similar faults i.e, external datasets.

- Minimum Validation Reconstruction Error Baseline (*val RE*): In anomaly detection settings where only healthy data is available, the autoencoder's reconstruction error on healthy validation data is often used as a proxy for model performance. Given that the autoencoder's objective is to minimize reconstruction error, selecting the model with the lowest reconstruction error provides a straightforward baseline for model selection.

- Mean Performance of All Trained Models Baseline (*Average*): Another simple approach is to measure the average performance of all models without hyperparameter optimization. This baseline represents a scenario where model selection is not tailored to any specific hyperparameter configuration and instead reflects a randomly chosen model's performance. Notably, this approach differs from ensembling, where multiple models' anomaly scores are combined. Here, we compute the mean performance across models solely as a benchmark to gauge the added value of model selection in anomaly detection.

- External datasets (*Ext-set*): Healthy/faulty data from external datasets has often been used to serve as a proxy for model selection. Here we simulate this exercise by assuming access to one of the faults and testing the model selection performance on the remaining fault datasets. In the table, *Ext-set1*, *Ext-set2*, *Ext-set3*, *Ext-set4*, *Ext-set5* refer to external datasets of Imbalance, Modulation, Whine, Wind and PWM faults respectively.

- Proxy anomalies (*Proposed*): The proposed method leverages engineered proxy-anomalies, which are fault-like perturbations applied to healthy validation samples. By incorporating proxy anomalies, an augmented validation dataset is created, that enables reliable model selection in the absence of real faulty data.

- Ideal model (*Ideal*): The ideal model selection scenario assumes access to labeled faulty data during model selection, allowing direct evaluation of model performance based on all faults. While this approach is not feasible in real life setting, it provides an upper-bound reference for evaluating the effectiveness of different model selection strategies.

## 6. Results and Discussion

This section presents the results of anomaly detection model using hyperparameters selected according to the proposed model selection methodology and its comparison with baseline techniques. Firstly, the results of different model selection strategies are evaluated to assess their effectiveness in anomaly detection across various fault types. We then showcase the performance of the optimal model identified through the methodology. The analysis highlights key findings, such as the model's strong performance on most fault types, and discusses challenges encountered with specific faults, like Modulation. The determination of appropriate thresholds, often viewed as an orthogonal problem to model selection, is beyond the scope of this paper [20; 69]. Thus, the Area Under the Receiver Operating Characteristic (AUROC) curve, which encapsulates model performance across a continuum of thresholds, has been selected as the evaluation metric.

### 6.1. Model selection

Table 6 shows the model architectures selected as a result of the hyperparameter tuning experiments, while Tables 7 and 8 show the performance of the selected models on the different fault datasets. This was done with 400 samples for training. Additional results with 800 and 200 training samples can be found in Appendix C. All experiments use 100 healthy validation samples, with an additional 100 augmented validation set samples in the case of proxy-anomaly based and external dataset based model selection strategies. Additionally, the test set is made up of 200 healthy and 200 faulty samples in all the experiments and all the faulty datasets. The dataset with all faults combined contains the same 200 healthy samples with faulty samples from all the five datasets combined (i.e 1000 faulty samples).

The results demonstrate that our approach is capable of selecting the best-performing anomaly detection models across all fault types. This highlights the reliability of the model selection strategy and its ability to generalize well across different fault conditions. The high test performance on these datasets confirms that effective anomaly detection is feasible even in the absence of real faulty data, provided that a well-structured model selection strategy is employed.

The Modulation fault proved significantly more challenging to detect then the others, with all trained models in the pool struggling to distinguish it from healthy samples. One possible reason is the small spectrogram region it affects, leading to a minimal impact on reconstruction error. However, this alone does not fully explain the issue, as the Whine fault, which also impacts a limited number of pixels, was still effectively detected. Another contributing factor is that the Modulation fault becomes audible at extremely low amplitudes, as shown in Table 2. In contrast, the Wind

**Table 6**

Selected hyperparameters using the different model selection criteria

| Model Selection Criteria | Model Selected | |
|---|---|---|
| | Number of Blocks | Number of Filters |
| Val RE | 1 | 32 |
| Ext-Set 1 | 7 | 4 |
| Ext-Set 2 | 5 | 4 |
| Ext-Set 3 | 6 | 8 |
| Ext-Set 4 | 5 | 8 |
| Ext-Set 5 | 7 | 16 |
| Proposed | 6 | 8 |
| Ideal | 6 | 8 |

**Table 7**

Test performance (AUROC) on five datasets with the different model selection criteria

| Model selection criteria | Dataset | | | | | |
|---|---|---|---|---|---|---|
| | Imbalance | Modulation | Whine | Wind | PWM | All faults |
| val RE | 0.53 ± 0.00 | 0.50 ± 0.00 | 0.51 ± 0.00 | 0.63 ± 0.00 | 0.58 ± 0.00 | 0.53 ± 0.00 |
| Average | 0.79 ± 0.17 | 0.53 ± 0.04 | 0.76 ± 0.2 | 0.91 ± 0.13 | 0.86 ± 0.18 | 0.74 ± 0.15 |
| Ext-set 1 | - | 0.54 ± 0.02 | 0.92 ± 0.01 | 1.00 ± 0.00 | 0.98 ± 0.01 | - |
| Ext-set 2 | 0.97 ± 0.03 | - | 0.92 ± 0.02 | 1.00 ± 0.00 | 0.98 ± 0.01 | - |
| Ext-set 3 | 0.89 ± 0.02 | 0.60 ± 0.04 | - | 1.00 ± 0.00 | 1.00 ± 0.00 | - |
| Ext-set 4 | 0.90 ± 0.01 | 0.56 ± 0.01 | 0.96 ± 0.01 | - | 1.00 ± 0.00 | - |
| Ext-set 5 | 0.92 ± 0.01 | 0.52 ± 0.01 | 0.94 ± 0.02 | 1.00 ± 0.00 | - | - |
| Proposed | 0.89 ± 0.02 | 0.60 ± 0.04 | 0.97 ± 0.01 | 1.00 ± 0.00 | 1.00 ± 0.00 | 0.88 ± 0.01 |
| Ideal | 0.89 ± 0.02 | 0.60 ± 0.04 | 0.97 ± 0.01 | 1.00 ± 0.00 | 1.00 ± 0.00 | 0.88 ± 0.01 |

**Table 8**

Correlation coefficients of different model selection metrics with test performance

| Model selection criteria | Dataset | | | | | |
|---|---|---|---|---|---|---|
| | Imbalance | Modulation | Whine | Wind | PWM | All faults |
| val_RE | 0.77 | 0.56 | 0.52 | 0.56 | 0.56 | 0.60 |
| Ext-set 1 | - | 0.64 | 0.90 | 0.85 | 0.91 | - |
| Ext-set 2 | 0.81 | - | 0.86 | 0.66 | 0.74 | - |
| Ext-set 3 | 0.91 | 0.72 | - | 0.84 | 0.94 | - |
| Ext-set 4 | 0.83 | 0.50 | 0.83 | - | 0.95 | - |
| Ext-set 5 | 0.91 | 0.60 | 0.94 | 0.96 | - | - |
| Proposed | 0.92 | 0.57 | 0.90 | 0.97 | 0.99 | 0.95 |

noise fault, which is also audible at low amplitudes, covers a much larger spectrogram region, leading to a stronger cumulative impact on reconstruction error. Additionally, the randomization introduced in healthy orders was set at 6 dB, which is relatively high compared to the fault amplitude itself, potentially masking subtle fault effects and making detection more difficult.

The proposed proxy-anomaly approach achieved strong model selection performance selecting models close to the ideal case. Similarly, external datasets also proved to be valuable indicators of model performance when available. Both strategies exhibited high correlation with test set performance, reinforcing their suitability as effective model selection methods. These findings highlight the robustness of proxy-anomalies and external datasets in guiding model selection for anomaly detection in data-scarce environments.

### 6.2. Anomaly detection model

To comprehensively evaluate the performance of the model selected according to the proposed criterion (model with 6 repeating blocks of 8 filters each), we analyze its behavior using multiple complementary visualizations. We first present a boxplot of reconstruction errors across all fault types to assess the model's capacity to distinguish healthy from faulty samples. Next, we explore the latent space distribution using t-distributed stochastic neighbor embedding (t-SNE), providing insight into how the model internally separates fault classes. We then examine pixel-level reconstruction error maps, which highlight spectral regions where the model struggles to reconstruct faulty inputs. Finally, saliency map analysis reveals the input regions that most strongly influence the model's anomaly detection decisions. Together, these visualizations offer a multifaceted understanding of the model's strengths and limitations in detecting and localizing faults.

### 6.3. Boxplot of reconstruction errors

Figure 11 shows a boxplot of the reconstruction errors for each fault type compared to healthy samples. The reconstruction errors for most fault classes are noticeably higher than those of the healthy data, highlighting the model's capacity to distinguish between normal and anomalous signals. However, variations within fault classes indicate differing levels of reconstruction difficulty.

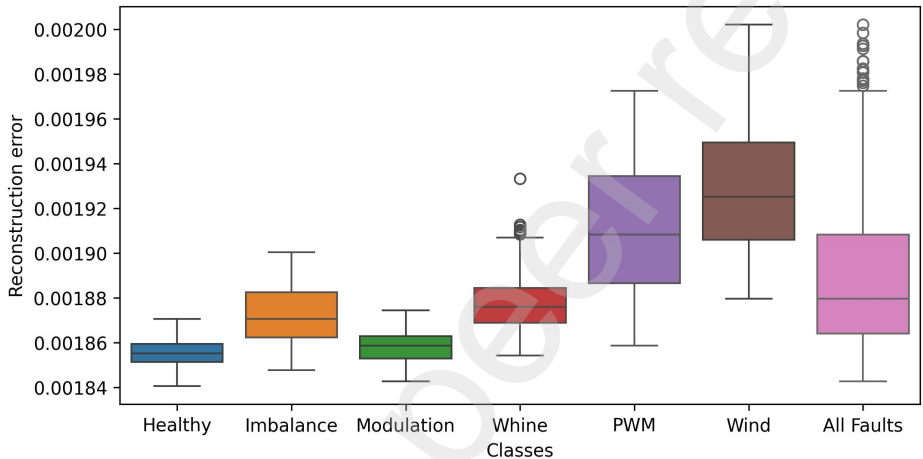

**Figure 11**: Boxplot of reconstruction errors of all the fault datasets as compared to the healthy samples

#### 6.3.1. Latent space visualization using t-SNE

To gain qualitative insight into the model's internal representation of different fault types, we visualize the latent space using t-SNE, applied to the encoded representation layer of the autoencoder. The resulting plot, shown in Figure 12, illustrates how the model organizes and separates various fault types in the latent space.

Prior to applying t-SNE, the latent representations were first reduced in dimensionality using Principal Component Analysis (PCA) to retain 90% of the original variance, resulting in 175 principal components. This initial step mitigates the sensitivity of t-SNE to high-dimensional noise and enhances its stability. The t-SNE algorithm was then used to project the PCA-compressed data into two dimensions for visualization. The parameters used for t-SNE were: perplexity = 300; learning rate = 100; number of iterations = 1000; initialization = 'random'; and cosine distance as the similarity metric.

The visualization reveals several meaningful patterns. The PWM and Wind noise faults show distinct clustering while lower order-related faults (e.g., Imbalance and Whine) form distinct parts of a sphere. In contrast, the Modulation fault cluster exhibits partial overlap with the healthy cluster, indicating a greater similarity in latent representation and potentially explaining lower detection performance for this fault type.

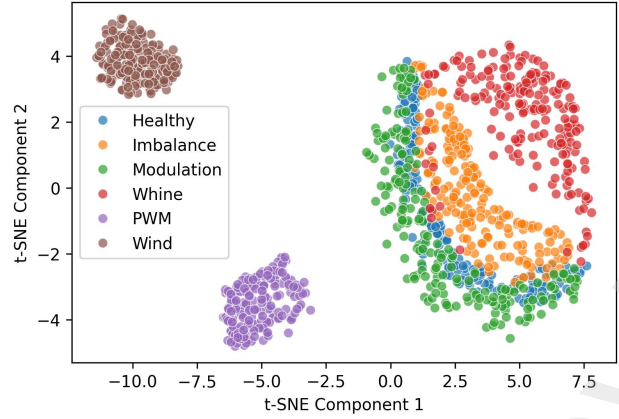

**Figure 12:** t-SNE visualization of the latent space representation for healthy and faulty data

### 6.3.2. Reconstruction error map analysis

To gain deeper insight into the model's reconstruction behavior across different fault types, we visualize the reconstruction error at a pixel level. Since the model is trained only on healthy data, deviations from the typical reconstruction pattern can indicate anomalous regions. To highlight these discrepancies, we apply the following postprocessing procedure to both healthy and faulty test samples:

- Computation of Reconstruction Error: Given a test spectrogram $x$ and its reconstruction $\hat{x}$, the pixel-wise reconstruction error is computed as:

$$E_{i,j} = (x_{i,j} - \hat{x}_{i,j})^2 \tag{9}$$

- Statistical Thresholding: The distribution of reconstruction errors is estimated from the training set, consisting of only healthy samples. Specifically, for each pixel $(i, j)$, we compute the mean $\mu_{i,j}$ and standard deviation $\sigma_{i,j}$ across all training samples:

$$\mu_{i,j} = \frac{1}{N} \sum_{n=1}^{N} E_{n,i,j}^{n} \tag{10}$$

$$\sigma_{i,j} = \sqrt{\frac{1}{N} \sum_{n=1}^{N} (E_{n,i,j}^{n} - \mu_{i,j})^2} \tag{11}$$

where $N$ is the number of training samples.

- Anomaly Highlighting: For each pixel in a test sample, we determine whether the reconstruction error significantly deviates from the normal range. Pixels falling outside three standard deviations from the mean are marked as anomalous:

$$M_{i,j} = \begin{cases} 1, & \text{if } E_{i,j} > \mu_{i,j} + 3\sigma_{i,j} \text{ or } E_{i,j} < \mu_{i,j} - 3\sigma_{i,j}, \\ 0, & \text{otherwise.} \end{cases} \tag{12}$$

This binary mask $M$ effectively highlights regions of high reconstruction error that deviate from the healthy distribution.

Figure 13 presents examples of reconstruction error (RE) anomaly maps across different fault types. This representation allows for a clearer identification of spectral regions most affected by each fault, providing insight into the model's ability to localize anomalies effectively. For comparison we also show the anomaly map for a healthy sound sample in Figure 13f.

For certain fault types, such as Whine, PWM, and Wind, the anomaly maps clearly highlight the faulty regions, demonstrating the model's capacity to pinpoint these anomalies with high accuracy. However, in the case of Imbalance, the highlighted region is less pronounced, making the fault somewhat less apparent. For Modulation, the model faces difficulty in distinguishing this fault from healthy data, resulting in an anomaly map that fails to highlight any distinct deviations.

The samples selected for plotting represent an average severity level of each fault type, providing a balanced and representative depiction of the model's performance across various levels of fault severity.

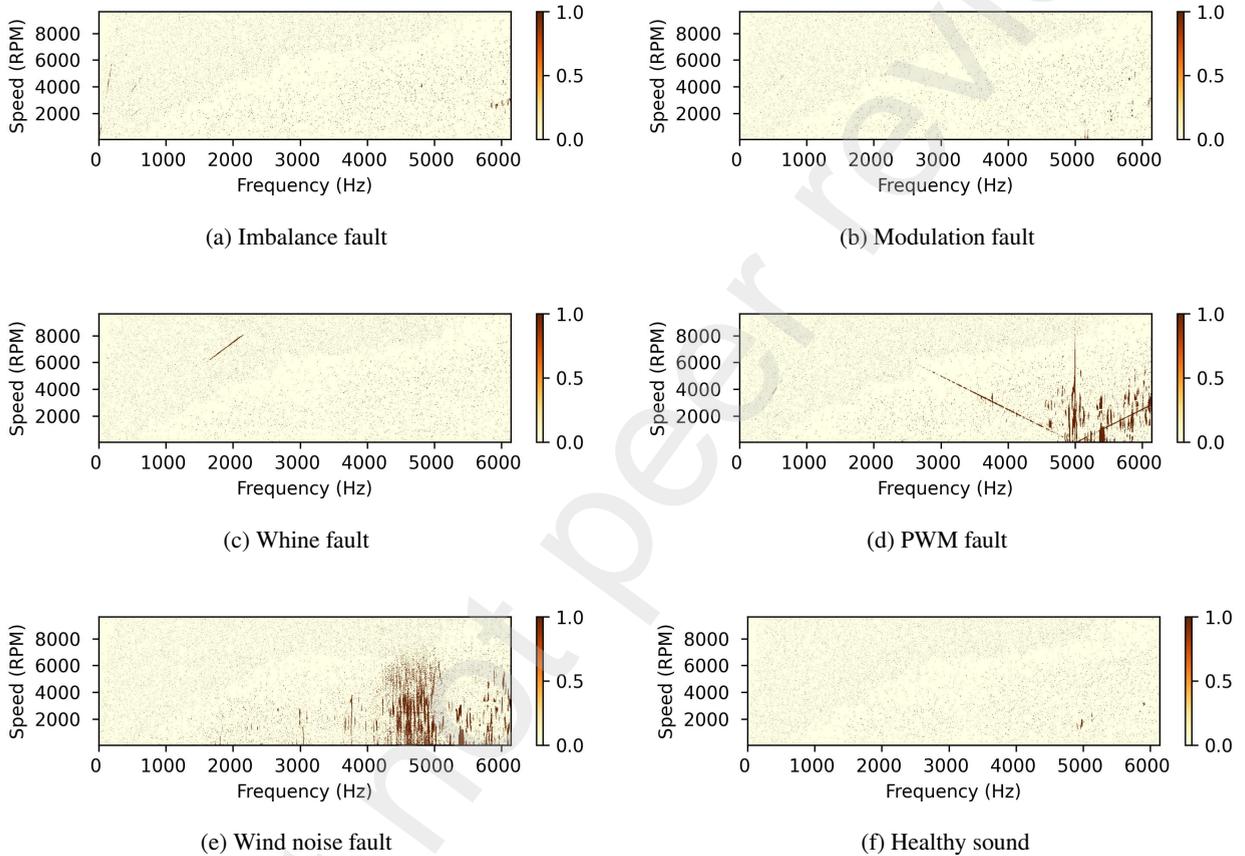

Figure 13: Reconstruction error maps for representative sounds from each fault category

### 6.3.3. Saliency map analysis

Saliency maps highlight areas in the spectrograms that contribute the most to the model's anomaly detection decision. Figure 14, presents the saliency maps for different fault types, illustrating the model's sensitivity to spectral regions most indicative of anomalies. They were computed by backpropagating the reconstruction loss and extracting the gradient of the loss with respect to the input spectrogram. The absolute value of the gradient is used to visualize the contribution of different features to the loss, highlighting areas the model considers important for reconstruction.

For all fault types, the highlighted regions in the saliency maps correspond to the expected fault locations. This alignment is particularly clear for faults like Whine and PWM. The healthy sample, as expected, shows minimal activation, confirming that the model does not falsely highlight regions in the absence of anomalies.

To ensure a representative evaluation, the selected samples correspond to mid-range severity levels for each fault type (same as in section 6.3.2).

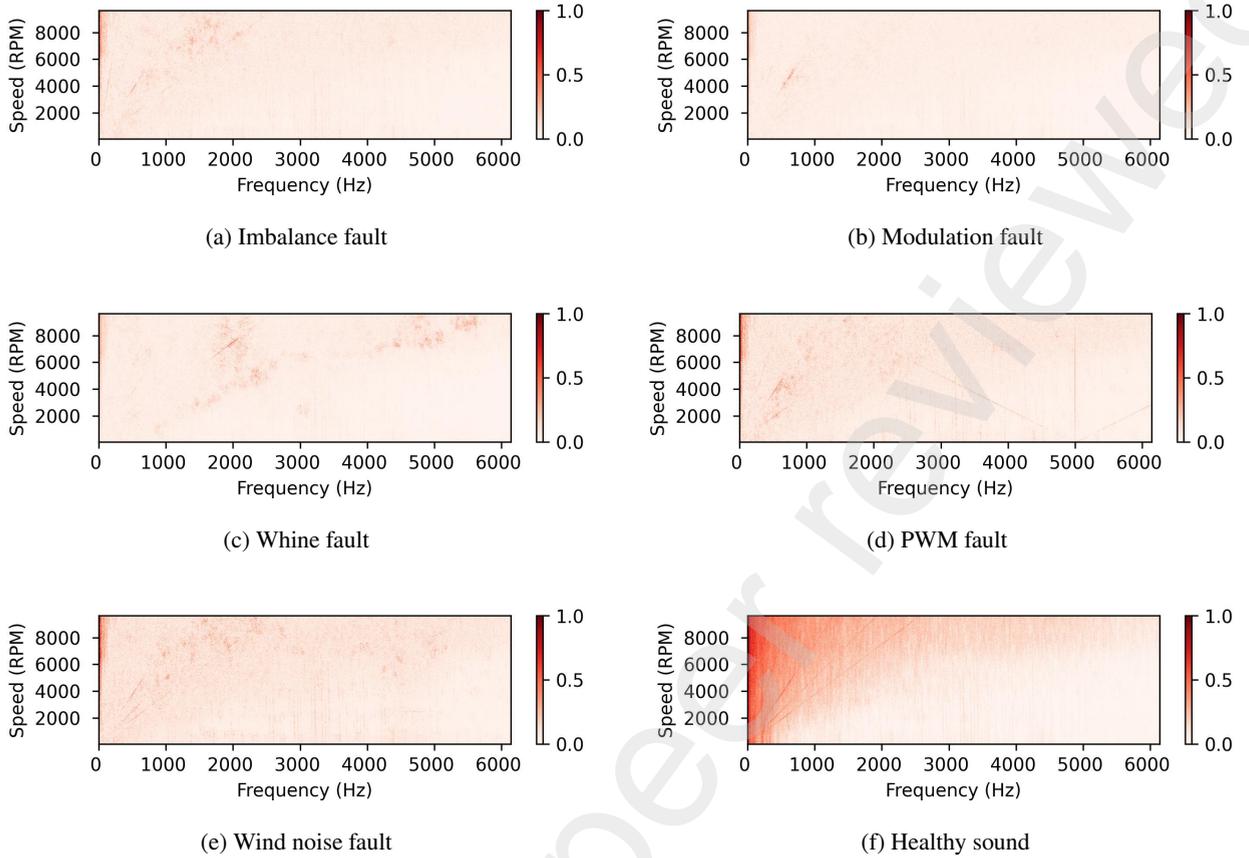

**Figure 14:** Saliency maps for representative sounds from each fault category

The results presented in this section provide a comprehensive evaluation of the proposed anomaly detection framework and its underlying model selection strategy. The analysis confirms that the proposed approach is highly effective in identifying optimal models across a wide range of fault types, achieving performance that is consistently close to the ideal benchmark. Notably, the use of proxy anomalies and external datasets has proven to be a robust strategy for guiding model selection in data-scarce settings. The complementary visual analyses, ranging from reconstruction errors and latent space embeddings to pixel-level error maps and saliency visualization, offer valuable insights into the model's strengths. Overall, these findings highlight the promise of the proposed approach for real-world anomaly detection tasks, especially in settings where labeled fault data is unavailable.

## 7. Conclusion

This study presented a domain-knowledge-driven framework for model selection in autoencoder-based anomaly detection in automotive interior sound, designed to address the critical challenge of limited faulty data during model development. By introducing proxy-anomalies i.e., structured perturbations applied to healthy spectrograms that mimic characteristics of real-world faults, the proposed approach enables effective model selection without requiring access to labeled faulty samples.

The framework was evaluated on a newly developed, high-fidelity dataset of healthy and faulty electric vehicle interior sounds covering five representative fault types: Imbalance, Modulation, Whine, Wind, and Pulse Width Modulation. Experimental results demonstrated that models selected using proxy-anomalies achieved performance trends closely aligned with those selected using real fault labels, significantly outperforming conventional unsupervised validation methods. This finding validates the utility of domain-informed proxy-anomalies as a reliable and interpretable solution for model selection in data-scarce settings.

Importantly, the curated dataset, synthesized using a sound quality equivalent model and validated through expert jury assessments, has been made publicly available, offering a valuable resource for advancing research in automotive acoustics and anomaly detection.

Overall, this work provides a scalable and generalizable strategy that addresses key limitations of traditional unsupervised validation approaches and enables more robust, data-efficient fault detection across diverse acoustic fault types. Future research could explore integrating proxy-anomalies directly into model training, refining perturbation strategies to increase their realism, and extending the framework to other acoustic monitoring applications. Incorporating adaptive learning mechanisms that tune proxy-anomaly generation to match real-world fault distributions also represents a promising avenue for further development.

*Acknowledgements* The authors gratefully acknowledge the support of the European Commission under the Marie Sklodowska Curie program through the ETN MOIRA project (GA 955681).

*Declaration of generative AI and AI-assisted technologies in the writing process* During the preparation of this work the authors used ChatGPT in order to improve readability of the work. After using this tool/service, the authors reviewed and edited the content as needed and take full responsibility for the content of the published article.

## A. Fault prevalence

Figure 15 shows the maximum area that each of the faults can affect. The half width of the order regions extracted is set to 15Hz.

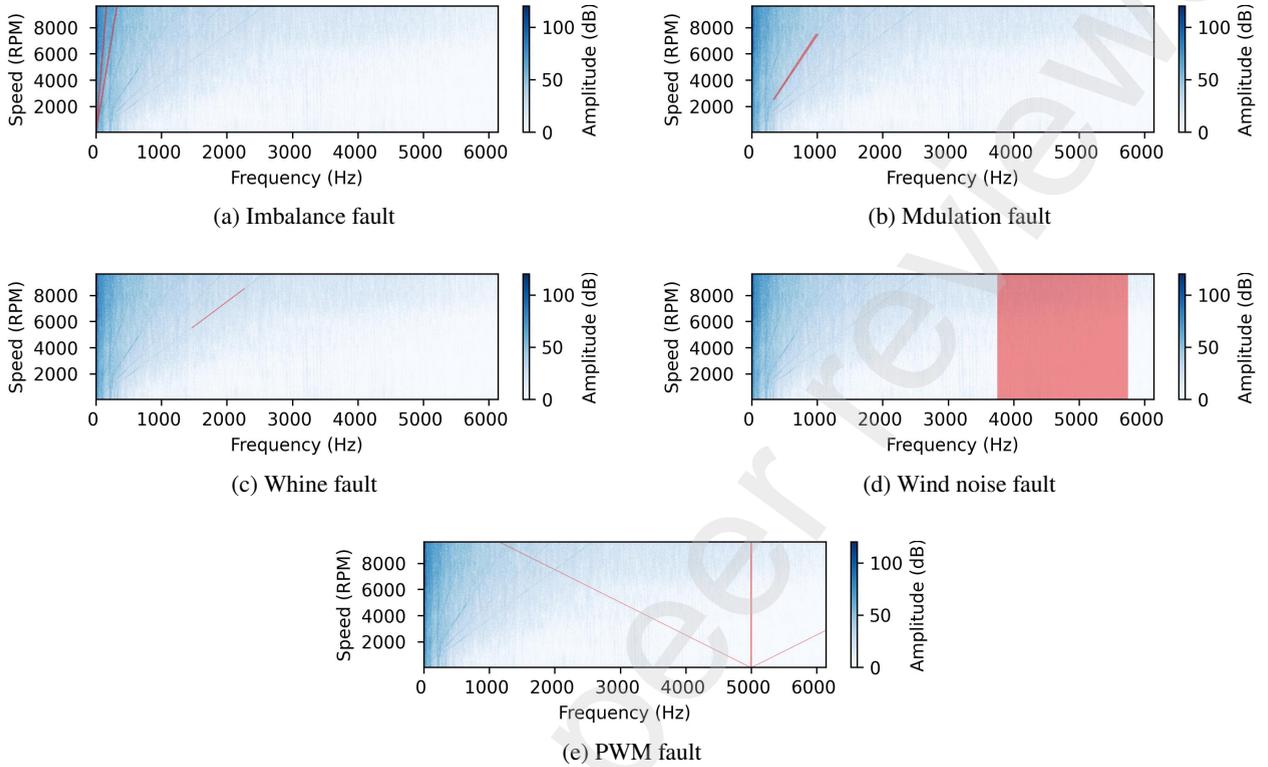

**Figure 15**: Fault prevalence maps show the maximum area affected by each fault category

## B. Proxy-anomaly examples

Figures 17, 18 and 19, show examples of spectrograms created with different types of perturbations of a healthy sound sample shown in Figure 16. The figures show the effect of random sampling on generation of the augmented validation set.

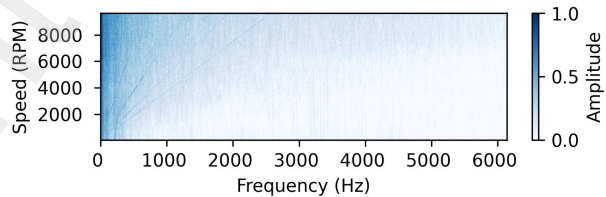

**Figure 16**: Original healthy data sample

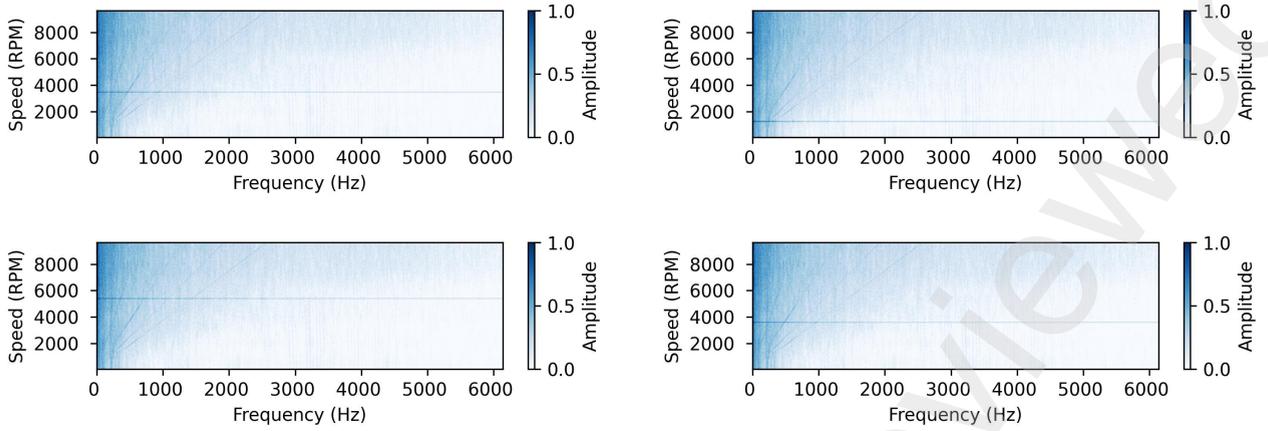

**Figure 17:** Proxy-anomalies created by the addition of a random RPM-line

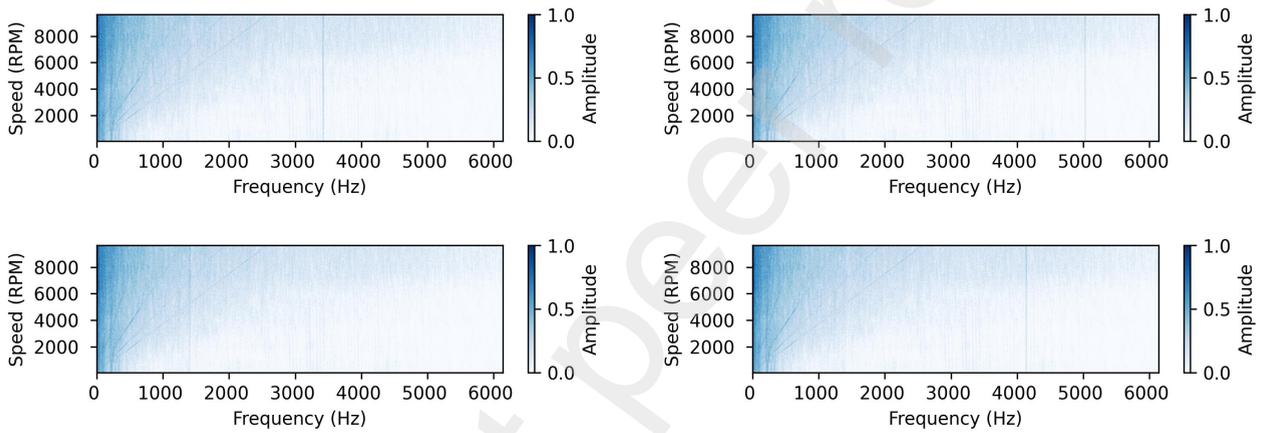

**Figure 18:** Proxy-anomalies created by the addition of a random Frequency-line

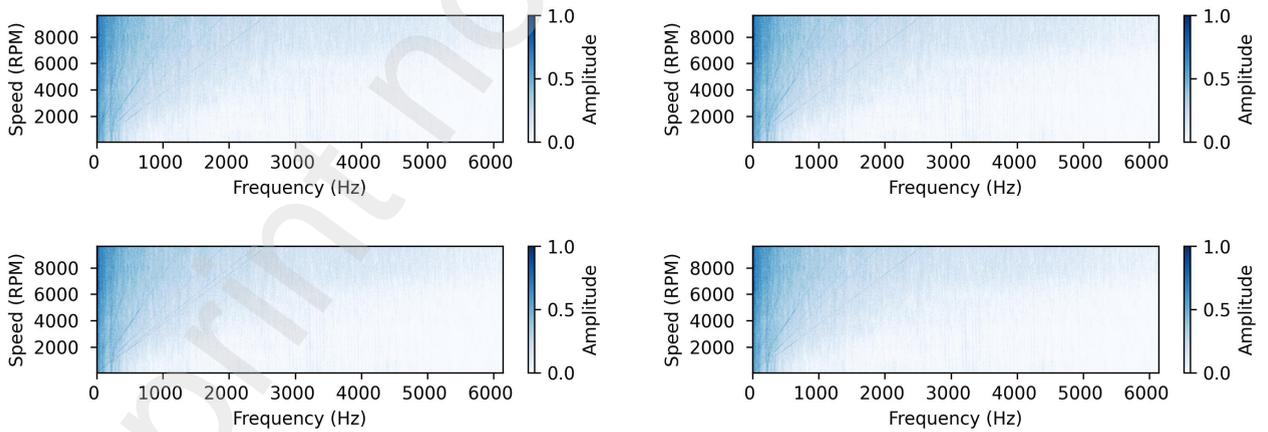

**Figure 19:** Proxy-anomalies created by the addition of a random Order-line

## C. Additional experiments

Tables 9, 10, 11, and 12 show the results of some additional experiments with 800 and 200 training samples.

**Table 9**
Test performance (AUROC) on five datasets with the different model selection criteria - 800 training samples

| Model selection criteria | Dataset | | | | | |
|---|---|---|---|---|---|---|
| | Imbalance | Modulation | Whine | Wind | PWM | All faults |
| val RE | 0.53 ± 0.00 | 0.50 ± 0.00 | 0.51 ± 0.00 | 0.64 ± 0.00 | 0.58 ± 0.00 | 0.52 ± 0.00 |
| Average | 0.75 ± 0.17 | 0.52 ± 0.04 | 0.75 ± 0.21 | 0.9 ± 0.14 | 0.85 ± 0.19 | 0.72 ± 0.16 |
| Ext-set 1 | - | 0.53 ± 0.02 | 0.97 ± 0.01 | 1.00 ± 0.00 | 1.00 ± 0.00 | - |
| Ext-set 2 | 0.89 ± 0.06 | - | 0.96 ± 0.01 | 1.00 ± 0.00 | 1.00 ± 0.00 | - |
| Ext-set 3 | 0.92 ± 0.02 | 0.58 ± 0.03 | - | 1.00 ± 0.00 | 1.00 ± 0.00 | - |
| Ext-set 4 | 0.87 ± 0.03 | 0.57 ± 0.02 | 0.94 ± 0.00 | - | 1.00 ± 0.00 | - |
| Ext-set 5 | 0.89 ± 0.01 | 0.58 ± 0.01 | 0.97 ± 0.00 | 1.00 ± 0.00 | - | - |
| Proposed | 0.92 ± 0.02 | 0.57 ± 0.03 | 0.96 ± 0.00 | 1.00 ± 0.00 | 1.00 ± 0.00 | 0.88 ± 0.01 |
| Ideal | 0.89 ± 0.06 | 0.63 ± 0.08 | 0.96 ± 0.01 | 1.00 ± 0.00 | 1.00 ± 0.00 | 0.89 ± 0.03 |

**Table 10**
Correlation coefficients of different model selection metrics with test performance - 800 training samples

| Model selection criteria | Dataset | | | | | |
|---|---|---|---|---|---|---|
| | Imbalance | Modulation | Whine | Wind | PWM | All faults |
| val RE | 0.85 | 0.67 | 0.84 | 0.94 | 0.91 | 0.88 |
| Ext-set 1 | 1.00 | 0.64 | 0.99 | 0.87 | 0.96 | 0.99 |
| Ext-set 2 | 0.85 | 0.94 | 0.88 | 0.65 | 0.75 | 0.87 |
| Ext-set 3 | 0.99 | 0.71 | 1.00 | 0.83 | 0.93 | 1.00 |
| Ext-set 4 | 0.84 | 0.48 | 0.82 | 1.00 | 0.95 | 0.86 |
| Ext-set 5 | 0.95 | 0.58 | 0.93 | 0.96 | 1.00 | 0.96 |
| Proposed | 0.93 | 0.56 | 0.91 | 0.98 | 0.99 | 0.94 |

**Table 11**
Test performance (AUROC) on five datasets with the different model selection criteria - 200 training samples

| Model selection criteria | Dataset | | | | | |
|---|---|---|---|---|---|---|
| | Imbalance | Modulation | Whine | Wind | PWM | All faults |
| val_RE | 0.55 ± 0.02 | 0.50 ± 0.00 | 0.51 ± 0.00 | 0.73 ± 0.00 | 0.58 ± 0.01 | 0.53 ± 0.00 |
| Average | 0.84 ± 0.17 | 0.52 ± 0.03 | 0.74 ± 0.18 | 0.85 ± 0.22 | 0.83 ± 0.2 | 0.73 ± 0.14 |
| Ext-set 1 | - | 0.53 ± 0.01 | 0.72 ± 0.06 | 0.75 ± 0.07 | 0.87 ± 0.04 | - |
| Ext-set 2 | 0.99 ± 0.01 | - | 0.92 ± 0.03 | 1.00 ± 0.00 | 0.98 ± 0.01 | - |
| Ext-set 3 | 0.89 ± 0.00 | 0.54 ± 0.01 | - | 1.00 ± 0.00 | 1.00 ± 0.00 | - |
| Ext-set 4 | 0.93 ± 0.02 | 0.54 ± 0.01 | 0.94 ± 0.01 | - | 0.99 ± 0.0 | - |
| Ext-set 5 | 0.92 ± 0.01 | 0.50 ± 0.01 | 0.91 ± 0.02 | 1.00 ± 0.00 | - | - |
| Proposed | 0.89 ± 0.00 | 0.54 ± 0.01 | 0.96 ± 0.01 | 1.00 ± 0.00 | 1.00 ± 0.00 | 0.87 ± 0.00 |
| Ideal | 0.99 ± 0.01 | 0.60 ± 0.04 | 0.92 ± 0.03 | 1.00 ± 0.00 | 0.98 ± 0.01 | 0.88 ± 0.02 |

**Table 12**
Correlation coefficients of different model selection metrics with test performance - 200 training samples

| Model selection criteria | Dataset | | | | | |
|---|---|---|---|---|---|---|
| | Imbalance | Modulation | Whine | Wind | PWM | All faults |
| val RE | -0.01 | -0.15 | -0.17 | -0.46 | -0.34 | -0.19 |
| Ext-set 1 | - | 0.68 | 0.86 | 0.52 | 0.83 | - |
| Ext-set 2 | 0.87 | - | 0.84 | 0.48 | 0.74 | - |
| Ext-set 3 | 0.88 | 0.67 | - | 0.73 | 0.92 | - |
| Ext-set 4 | 0.51 | 0.42 | 0.76 | - | 0.89 | - |
| Ext-set 5 | 0.82 | 0.60 | 0.93 | 0.87 | - | - |
| Proposed | 0.88 | 0.61 | 0.90 | 0.74 | 0.96 | 0.94 |